\documentclass[preprint,12pt]{elsarticle}

\usepackage{lineno,hyperref}
\modulolinenumbers[5]
\usepackage{graphicx}
\usepackage{amssymb}
\usepackage{tabularx}
\journal{Sol. Energy Mater. Sol. cells}

\begin{document}

\begin{frontmatter}

\title{Transition metals doped CuAlSe$_2$ for promising intermediate band materials}

\author[1]{Tingting Wang}
\author[1]{Xiaoguang Li}
\author[1]{Wenjie Li}
\author[2]{Li Huang}
\author[1]{Ya Cheng}
\author[1]{Jun Cui}
\author[1]{Hailin Luo}
\author[1]{Guohua Zhong\corref{cor1}}
\cortext[cor1]{Corresponding author}
\ead{gh.zhong@siat.ac.cn}

\author[1]{Chunlei Yang\corref{cor2}}
\cortext[cor2]{Principal corresponding author}
\ead{cl.yang@siat.ac.cn}

\address[1]{Center for Photovoltaic Solar Energy, Shenzhen Institutes of Advanced Technology, Chinese Academy of Sciences, Shenzhen, 518055, China}
\address[2]{Department of Physics, South University of Science and Technology of China, Shenzhen, 518055, China}

\begin{abstract}
Introducing an isolated intermediate band (IB) into a wide band gap semiconductor can potentially improve the optical absorption of the material beyond the Shockley-Queisser limitation for solar cells. Here, we present a systematic study of the thermodynamic stability, electronic structures, and optical properties of transition metals ($M=$ Ti, V, and Fe) doped CuAlSe$_2$ for potential IB thin film solar cells, by adopting the first-principles calculation based on the hybrid functional method. We found from chemical potential analysis that for all dopants considered, the stable doped phase only exits when the Al atom is substituted. More importantly, with this substitution, the IB feature is determined by $3d$ electronic nature of $M^{3+}$ ion, and the electronic configuration of $3d^1$ can drive a optimum IB that possesses half-filled character and suitable subbandgap from valence band or conduction band. We further show that Ti-doped CuAlSe$_2$ is the more promising candidate for IB materials since the resulted IB in it is half filled and extra absorption peaks occurs in the optical spectrum accompanied with a largely enhanced light absorption intensity. The result offers a understanding for IB induced by transition metals into CuAlSe$_2$ and is significant to fabricate the related IB materials.
\end{abstract}

\begin{keyword}
Intermediate band, Intermediate band solar cell, CuAlSe$_2$, first-principles calculation
\end{keyword}

\end{frontmatter}

\section{Introduction}
The conversion efficiency is one of the most important factors to estimate the performance of photovoltaic solar cells. At present, many ideas to increase the efficiency of solar cell have been developed including the third generation solar cells. As one of the third generation solar cells, intermediate band solar cell (IBSC) is believed to have bright prospects. For traditional photovoltaic solar cells, electrons are directly excited from the valence band (VB) to the conduction band (CB) by absorbing photons. For IBSC, the concept is to introduce an isolated intermediate band (IB) in the main band gap of the host semiconductor to build a three-photon absorption process such that electrons can be excited not only from the VB to the CB but also from the VB to the IB and from the IB to the CB. As a result, the upper limiting efficiency of IBSC was predicted to be as high as 63.1\%, greater than the single-junction solar cell (40.7\%).\cite{ref1,ref2} By further increasing the number of IBs, the efficiency would reach 86.8\%.\cite{ref3,ref4} Theoretical and experimental reports have verified that IB materials could effectively increase the optical absorption.\cite{ref5,ref6,ref7} Numerous efforts have been made to implement IBSCs, including through quantum dots and the insertion of appropriate impurities into the bulk host semiconductor.\cite{ref5,ref6,ref7,ref8,ref9,ref10,ref11,ref12,ref13,ref14,ref15,ref16,ref17,ref18}

From previous studies, the bulk IB material is easier to fabricate than quantum dots and has a stronger absorption because of the higher density of the IB states.\cite{ref18} Among bulk IB host materials, Cu-based chalcopyrite compounds are the important candidate host semiconductors for IBSCs with the high conversion efficiency of 46.7\%.\cite{ref19} When the value of band gap is in the optimum region of $2.2-2.8$ eV with the IB located in the region of $0.8-1.1$ eV from VB, the photovoltaic energy conversion is becoming stronger. Hence, CuGaS$_2$ and CuAlSe$_2$ are viewed as the most promising IB host materials because their band gaps are 2.45 eV and 2.67 eV, respectively. In the past decade, many theoretical studies, based on the first-principles calculations, have been made to design and understand the bulk CuGaS$_2$ IB materials.\cite{ref20,ref21} Elements of the $3d$ transition metals and group IVA have been identified as viable candidates to act as substitutes in the cation sites of the CuGaS$_2$ chalcopyrite hosts.\cite{ref14,ref20,ref21,ref22,ref23,ref24,ref25,ref26,ref27,ref28} In particular, Ti\cite{ref29}, Sn\cite{ref30}, and Fe\cite{ref31} doped CuGaS$_2$ IB semiconductors have been successfully synthesized and display the enhancement of absorption coefficient. However, the photovoltaic energy conversion does not meet the expectations of higher efficiency. More efforts are needed in the future work.

With respect to CuAlSe$_2$ IB materials, the band gap of the host semiconductor is about 2.6 eV. After introducing a IB located at 1.01 eV from VB, the photovoltaic energy conversion will realize 46\%.\cite{ref19} Comparing to CuGaS$_2$, perfect CuAlSe$_2$ has much higher adsorption coefficient in the visible region.\cite{ref28,ref32} Nevertheless, few studies on CuAlSe$_2$ IB materials are reported. For transition metals, we know that introducing local $3d$ electrons into the wide band gap semiconductor often induces the impurity bands in the main band gap of the host. Considering the advantage of $3d$ electrons of transition metals and selecting three typical elements Ti, V, and Fe, in this work, we therefore employ density functional theory based on the hybrid density functional method to investigate the transition metals $M$-doped CuAlSe$_2$. The effects of doping concentrations and positions are examined. It is our aim to explore the thermodynamic stability induced by doping, understand the forming mechanism of IB, and predict the promising candidates materials for IBSC. This is significant to fabricate the CuAlSe$_2$ IB materials and develop the related IBSCs.

\section{Computational Details}
Density functional theory (DFT) calculations have been carried out on a representative structure which consists on a supercell derived from the parent body-centered tetragonal CuAlSe$_2$ structure as shown Fig. 1(a). Different concentrations of the proposed dopants inside the host semiconductor were computed. In order to obtain the expectative doping concentration,  calculations for supercells containing 16, 32, 64, and 128 atoms were carried out and compared. The desired structures were achieved by substituting $M$ atom for Cu, for Al, and for Se atom respectively inside a CuAlSe$_2$ supercell. The configurations were fully optimized using a conjugate-gradient algorithm. The Monkhorst-Pack $k$-point grids are generated according to the specified $k$-point separation 0.02 {\AA}$^{-1}$ and the convergence thresholds are set as $10^{-6}$ eV in energy and 0.005 eV/{\AA} in force.

All calculations are based on the projector augmented wave method (PAW)\cite{ref33} with a cutoff energy of 400 eV as implemented in the Vienna \emph{ab} initio simulation package (VASP)\cite{ref34}. In the standard DFT, the generalized gradient approximation (GGA) of Perdew-Burke-Ernzerhof (PBE) version\cite{ref35} is adopted to describe the electronic exchange-correlation (XC) interactions. However, the standard DFT usually leads to erroneous descriptions for some real systems such as underestimating the band gap. One way of overcoming this deficiency is to use Heyd-Scuseria-Ernzerhof (HSE) hybrid functional, where a part of the nonlocal Hartree-Fock (HF) type exchange is admixed with a semilocal XC functional, to give the following expression\cite{ref36}:
\begin{eqnarray}
E^{HSE}_{xc}=&&\alpha E^{HF,SR}_{x}(\nu)+(1-\alpha)E^{PBE,SR}_{x}(\nu)\nonumber \\
&&+E^{PBE,LR}_{x}(\nu)+E^{PBE}_{c},
\end{eqnarray}
where $\alpha$ is the mixing coefficient and $\nu$ is the screening parameter that controls the decomposition of the Coulomb kernel into short-range (SR) and long-range (LR) exchange contributions. In this calculation, the HSE exchange-correlation functional is set as 35\% mixing of screened HF exchange to PBE functional, namely $\alpha=0.35$, which can reproduce the experimental band gap of CuAlSe$_2$. Noticeably, Spin polarization is studied due to the existence of a magnetic $M$ atom.

\section{Results and discussion}

CuAlSe$_2$ crystallizes in a chalcopyrite structure with a space-group of $I\bar{4}2d$. As shown in Fig. 1(a), each atom in this structure is fourfold coordinated. Namely, each Se atom is coordinated with two Al and two Cu atoms, and each cation is coordinated with four Se ions. The optimized lattice constants of CuAlSe$_2$ are $a = 5.654$ {\AA} and $c=11.147$ {\AA} according with the experimental values. The calculated band gap is only 0.86 eV within the GGA method, while it reaches 2.6 eV based on HSE functional ($\nu=0.35$), which is in good agreement with the experimental value of 2.67 eV.\cite{ref37} From the electronic characteristics near the Fermi level shown in Fig. 1(b), the VB is mainly formed by Cu-$3d$ and Se-$4p$ electronic states, while the CB mainly results from Al-$3s$ and Se-$4s4p$ states.

To examine the probability of the formation of doped compounds, we have calculated the formation energy of doped system using the following as
\begin{eqnarray}
E_{f}=&&E_{SC}(doped)-E_{SC}(host)+\sum_{i}\Delta n_i(E_i+\mu_i)\nonumber \\
=&&\Delta H_{f}+\sum_{i}\Delta n_i\mu_i,
\end{eqnarray}
where $E_{SC}(doped)$ and $E_{SC}(host)$ is the total energy of the doped and host perfect supercells, respectively. $E_i$ is the total energy of the component element $i$ in its pure phase, which has been shown in Table I. $\Delta n_i$ is the number of exchanged atom $i$ between the host and the doped system ($\Delta n_i$ means that the atom $i$ moves into the host). $\mu_i$ is the chemical potential referenced to the total energy $E_i$ of the pure element phase for the $i$ species, satisfying $\mu_{i}\leq0$. $\mu_{i}=0$ represents the limit where the element is so rich that its pure phase can form. $\mu_{i}<0$ means that the formation of the compound is favorable rather than the pure elemental phase. $\Delta H_{f}$ is the formation enthalpy of doped system related to formation energy. $E_{f}$ represents the chemical reaction ongoing, while $E_{f}$ indicates that the thermodynamic balance is achieved between the host and doping compounds.

To analyze the thermodynamic stability of doped systems, we firstly start from the growth of the host CuAlSe$_2$. To obtain the CuAlSe$_2$ phase under a certain chemical environment, we have a balanced system with $E_{f}=0$ , and the chemical potentials of Cu, Al, and Se satisfy the equation
\begin{equation}
\mu_{Cu}+\mu_{Al}+2\mu_{Se}=\Delta H_{f}(CuAlSe_2)=-4.24 ~eV,
\end{equation}
where the formation enthalpy $\Delta H_{f}$(CuAlSe$_2$) can be obtained from our first-principles calculations. At the same time, to avoid the secondary phases Cu$_2$Se and Al$_2$Se$_3$, the chemical potentials satisfy the following relations
\begin{equation}
2\mu_{Cu}+\mu_{Se} < \Delta H_{f}(Cu_2Se)=-0.78 ~eV,
\end{equation}
and
\begin{equation}
2\mu_{Al}+3\mu_{Se}<\Delta H_{f}(Al_2Se_3)=-6.79 ~eV.
\end{equation}
By considering the above constraints Eqs. (3-5), we can draw the chemical potential range of stable CuAlSe$_2$ phase excluding the secondary phases of Cu$_2$Se and Al$_2$Se$_3$. Based on the analysis, as long as we know the formation enthalpy of doped systems and the chemical potential of different species $i$, the formation energy of the doped systems can be evaluated from Eq. (2). The chemical potential of different species $i$ has been listed Table I. We have also calculated the formation enthalpy of different doped systems and show in Table II. Considering the doped system with the transition metal $M$ occupying on the Cu site, namely $M_{Cu}$, we should control the chemical potentials to satisfy
\begin{equation}
\Delta H_{f}(M_{Cu})+\mu_{Cu}-\mu_{M}<0.
\end{equation}
Similarly, for the $M$ respectively occupying on the Al and Se site, $M_{Al}$ and $M_{Se}$, the chemical potentials must respectively satisfy
\begin{equation}
\Delta H_{f}(M_{Al})+\mu_{Al}-\mu_{M}<0
\end{equation}
and
\begin{equation}
\Delta H_{f}(M_{Se})+\mu_{Se}-\mu_{M}<0.
\end{equation}

By applying the above constraints in Eqs. (3-8), we can obtain the chemical potential range for the doped CuAlSe$_2$ system at $\mu_{M}=0$ (namely under $M$-rich condition). As shown in Fig. 2, the direction faced the origin satisfies the chemical potential constraints when substituting for Cu atom (plotted with solid line), while the direction deviated from the origin satisfies the chemical potential constraints when substituting for Al or Se atom (plotted with dashed line). When $\mu_{M}<0$ (namely under $M$-poor condition), the solid line will further shift toward to the origin, while the dashed lines further deviate from the origin. From the chemical potential range shown in Fig. 2(a), therefore, the doped systems of Ti$_{Cu}$ and Ti$_{Se}$ is difficult to grow. The grey region shown in Fig. 2(a) implies the chemical potential range to be able to obtain the doped system of Ti$_{Al}$. Lowering the chemical potential of Ti, namely representing the Ti-poor condition, the line determined by Ti$_{Al}$ will shift toward to the left, which results in the area of the grey region in Fig. 2(a) reducing. As a result, the Ti$_{Al}$ can be obtained when $\Delta H_{f}$(Ti$_{Al})-3.85\leq\mu_{Ti}\leq0$ eV, where $\Delta H_{f}$(Ti$_{Al})$ can be obtained from our first-principles calculations, such as the values shown in Table II. The doping concentration is just corresponding to the variation of $\mu_{Ti}$. For our considered several doping concentrations, as shown in Fig. 2(a), the phase boundary of Ti$_{Se}$ is sensitive to the doping content.

With regarded to the substitution V atom for Cu or Se atom, as shown in Fig. 2(b), the doped systems of both V$_{Cu}$ and V$_{Se}$ are difficult to grow from the chemical potential constrains. For the substitution V atom for Al atom, the line determined by V$_{Al}$ shifts toward to the left comparing with Ti$_{Al}$. The grey region shown in Fig. 2(b) marks out the range of chemical potential to grow the V$_{Al}$ system. However, the area of grey region in Fig. 2(b) slightly reduces comparing with that in Fig. 2(a). The V$_{Al}$ can be stabilized in the range of $\Delta H_{f}$(V$_{Al})-3.85\leq\mu_{V}\leq0$ eV. For the Fe doping situation, the Fe$_{Se}$ is impracticable, while the Fe$_{Cu}$ can be obtained in a very small chemical potential region. When substituting Fe atom for Al atom, the Fe$_{Al}$ can be grown in a region of chemical potential as the grey area shown in Fig. 2(c). Namely, the doping at Al site is available when the range of Fe chemical potential satisfies $\Delta H_{f}$(Fe$_{Al})-3.85\leq\mu_{Fe}\leq0$ eV. Thus far, we have cleared that the stable doped phase can be obtained only when substituting $M$ for Al atom, which is according with transition metal doped CuGaS$_2$. Additionally, when presenting the chemical potential range of achieving in growth, we assume the constraint of Cu-rich ($\mu_{Cu}=0$). However, the experimental fabrication often achieves under the Cu-poor situation ($\mu_{Cu}<0$). For $\mu_{Cu}<0$, the area of the grey region in Fig. 2 will decrease. Combining with Eqs. (3-8), to obtain the doped phase, we know that the chemical potential of Cu must satisfy $-0.85<\mu_{Cu}\leq0$ eV.

After obtaining the stable $M$-doped CuAlSe$_2$, we further investigate the electronic structures to evaluate the feasibility using as IB materials. View from the previous study\cite{ref19}, to achieve the high conversion efficiency, the subbangap formed by IB from VB or CB should be in the range of $0.7-1.2$ eV, or near to this range. A narrow IB also leads to a high effective mass and a low carrier mobility that makes carrier transport within the band difficult. Thus IB with the finite width is perfect\cite{ref38}. Importantly, the IB need be half-filled with electrons since both electrons and empty states are required to supply electrons to the CB as well as accepting them from the VB\cite{ref18}. To examine whether the impurity bands induced by $M$ doping satisfy these requirements above, we have calculated the electronic density of states (DOS) of $M$-doped CuAlSe$_2$ for different doping situations. Figure 3(a) shows the DOSs for Ti doping concentrations of 25\% and 6.25\%, respectively. Clear IB localized around the Fermi level is observed in these two doped CuAlSe$_2$ systems with substituting Ti atom for Al atom. As the dopant content reaching 25\%, the IB is 1.4 eV from the VB maximum (VBM) and 0.7 eV from the CB minimum (CBM). The width of IB is about 0.5 eV. Lowering the doping concentration to 6.25\%, the subbandgaps correspondingly become to 1.6 eV and 1.0 eV. The IB position in the Ti-doped system is suitable for promising IB materials as proposed by Mart\'{\i} \emph{et al}\cite{ref19}. With the increase of doping content as shown in Fig. 3(a), the electronic states to form IB increase and result in the extending of IB, which will suppress the Shockley-Read-Hall recombination\cite{ref28,ref39}. Seen from the DOSs of V-doped CuAlSe$_2$ at two concentration levels of 25\% and 6.25\% shown in Fig. 3(b), however, no half-filled IB is observed around the Fermi level. The impurity bands are far away from the Fermi level and close to the VBM or CBM which indicates the subbandgap is very small. Furthermore, we can not observe the half-filled IB in Fe-doped CuAlSe$_2$ systems either. As the DOSs of two doping concentrations shown in Fig. 3(c), the IB is completely empty though it has the isolated feature. As a result, from the aspect of electronic characteristics of IBs, both V-doping and Fe-doping are not good candidates, though the substitutions of V or Fe atom for Al atom satisfy the conditions of thermodynamic stability.

The IB feature is related to the electronic configuration of transition metallic ion. In chalcopyrite CuAlSe$_2$, four nearest neighboring Se atoms around a Al atom form a tetrahedral crystal field. The $M$ ion substitutes the Al atom in CualSe$_2$ and therefore is located in a tetrahedral crystal field environment. Based on the crystal field theory, the fivefold degenerated $3d$ orbital of $M$ will split into two main manifolds: lower twofold degenerate $e_g$ states and upper threefold degenerate $t_{2g}$ states. Considering the electronic spin polarization on $3d$ orbitals, $e_g^{2\uparrow}t_{2g}^{3\uparrow}$, namely $d^{5\uparrow}$, represents the five $3d$ electrons with the majority-spin, while $e_g^{2\downarrow}t_{2g}^{3\downarrow}$ ($d^{5\downarrow}$) corresponding to the minority-spin situation. Substituting for Al atom in CuAlSe$_2$, Ti, V, and Fe ions exhibit the valence states of Ti$^{3+}$, V$^{3+}$, and Fe$^{3+}$, respectively. The orbital splitting in the tetrahedral crystal field combined with the DOS of $M$-$3d$ leads to the electronic configurations of Ti$^{3+}$-$3d^1$-$e_g^{1\uparrow}$, V$^{3+}$-$3d^2$-$e_g^{2\uparrow}$, and Fe$^{3+}$-$3d^5$-$e_g^{2\uparrow}t_{2g}^{3\uparrow}$. $M^{3+}$ ion is at high-spin state with all $3d$ electrons being spin up. For the substitution Ti atom for Al atom shown in Fig. 3(a), the IB nature origins from the half filling of $e_g^{\uparrow}$ band, due to the single $3d$ electron in Ti$^{3+}$. The coupling between $e_g^{\uparrow}$ band and VBs is very weak. So there is visible gap between them. For the substitution V atom for Al atom, the number of $3d$ electrons increases to two. However, as shown in Fig. 3(b), these two $3d$ electrons full fill the $e_g^{\uparrow}$ band and lower its energy, which results that the full-filled $e_g^{\uparrow}$ band strongly couples with VB. The empty $3d$ bands made up of $t_{2g}^{\uparrow}$ and $e_g^{\downarrow}t_{2g}^{\downarrow}$ are pushed up to the CB. This is reason that the favorable IB can not be observed in V-doped CuAlSe$_2$. In Fe doping case, all of five $3d$ electrons exhibit the majority spin nature, namely $e_g^{2\uparrow}t_{2g}^{3\uparrow}$. The full-filled $e_g^{\uparrow}t_{2g}^{\uparrow}$ bands strongly hybridize with the VBs of the host below the Fermi level. The $e_g^{\downarrow}t_{2g}^{\downarrow}$ bands are completely empty and form the isolated IBs in the main band gap of CuAlSe$_2$. This is just the origin that no half-filled IBs is observed in Fe-doped CuAlSe$_2$.

Along the analysis for three doping elements above and extending other $3d$ transition metals such as Cr, Mn, Co, and Ni, the $3d$ electronic configuration is respectively Cr$^{3+}$-$3d^{3}$-$e_g^{2\uparrow}t_{2g}^{1\uparrow}$, Mn$^{3+}$-$3d^{4}$-$e_g^{2\uparrow}t_{2g}^{2\uparrow}$, Co$^{3+}$-$3d^{6}$-$e_g^{2\uparrow}t_{2g}^{3\uparrow}e_g^{1\downarrow}$, and Ni$^{3+}$-$3d^{7}$-$e_g^{2\uparrow}t_{2g}^{3\uparrow}e_g^{2\downarrow}$ under substituting for Al atom in CuAlSe$_2$. For Ni$^{3+}$, we find that the orbitals are all full occupied without the half-filled band around the Fermi level. Seen from the electronic configurations, the half-filled or unfilled band is existent, $t_{2g}^{1\uparrow}$ band for Cr$^{3+}$, $t_{2g}^{2\uparrow}$ band for Mn$^{3+}$, and $e_g^{1\downarrow}$ band for Co$^{3+}$, which can be viewed as IBs. However, the full-occupied band of $e_g^{2\uparrow}$ or $t_{2g}^{3\uparrow}$ that exists will strongly couple or hybridize with the VBs of the host. As a result, the subbandgap between VB and IB will reduce or disappear. Combining this analysis with three typical doping elements of Ti, V, and Fe, therefore, we conclude that the IB characteristic is determined by the electronic nature of $M^{3+}$ ion. We further infer that $M^{3+}$ ion with the electronic configuration of $d^1$ can induce the optimum IB in wide band gap semiconductors. Just this, Ti is the most promising candidate.

Adopted the method of Gajdo\v{s} \emph{et al.}\cite{ref40}, the calculated optical properties further confirm the promising of Ti-doped CuAlSe$_2$ IB materials. Figure 4 shows the absorption coefficient $\alpha$ of Ti-doped CuAlSe$_2$ in visible light region, comparing with the host. After doping, we can observe a greatly enhanced light absorption intensity and additional absorption peaks in the range of $0.7-2.2$ eV, which is induced by the IB. Especially, the 25\% doping concentration case (Cu$_4$Al$_3$TiSe$_8$) visibly displays the three-photon absorption process. As shown in Fig. 4, peak 1 implies the electronic transition from IB to CB, and peak 2 represents the electronic transition from VB to IB, while peak 3 results from the electronic transition from VB to CB. The optical result further supports that Ti is the promising candidate to achieving the IB material.

\section{Conclusions}
To explore and design novel structural solar cell materials with higher conversion efficiency, we have investigated the feasibility of $M$ (Ti, V, and Fe) doped CuAlSe$_2$ as IBSC material from thermodynamic stability, electronic structures, and optical properties, using the first-principles calculation based on HSE hybrid functional. Based on the chemical potential analysis, we point out that the stable $M$-doped CuAlSe$_2$ only exists when substituting for Al atom, and present the chemical potential conditions to growth. Seen from electronic structures, the Ti substitution results in the optimum IB feature, such as half-filled, favorable subbandgaps of 1.4 eV from VB to IB and 0.7 eV from IB to CB, as well limit width of IB for 25\% doping concentration. However, the impurity band induced by the V doping adjoins the VB or CB, while the IB formed by the Fe doping is completely empty. Stable V- and Fe-doped phase can not result in the half-filled IB. We therefore conclude that the $M^{3+}$ with the electronic configuration of $3d^1$ can drive the optimum IB in the wide bandgap semiconductor, namely the IB feature origins from the $3d$ electronic nature into the host. Combining the analysis above, Ti is considered to be the more promising candidate. The calculated absorption spectra of Ti-doped CuAlSe$_2$ illustrate three-photons absorption process. Two additional absorption peaks are formed in the band gap region of the host. And the absorption coefficient of doped systems in visible light region is obviously stronger than that of the host material. This study is helpful for the experimental research of IBSC materials from the doping stability and the forming mechanism of IB.

\section*{Acknowledgments}
The work was supported by the National Major Science Research Program of China under Grant no. 2012CB933700, the Natural Science Foundation of China (Grant nos. 61274093, 61574157, 11274335, 11504398, 51302303, 51474132, and 11404160), and the Basic Research Program of Shenzhen (Grant nos. JCYJ20150529143500956, JCYJ20140901003939002, JCYJ20150521144320993, JCYJ20140417113430725, and JCYJ20150401145529035).


\newpage

\begin{table}
\caption{\label{tab:table1}
Total energy per atom (eV) of the pure element phase.}
\begin{tabular}{cc}
\hline
Phase                          & $E_{i}$  \\ \hline
Cu face-centered cubic bulk    & -3.72 \\
Al face-centered cubic bulk    & -3.74 \\
Se$_2$ molecule                & -2.63 \\
Ti hexagonal close-packed bulk & -7.76 \\
V body-centered cubic bulk     & -8.94 \\
Fe body-centered cubic bulk    & -7.83 \\ \hline
\end{tabular}
\end{table}

\begin{table}
\caption{\label{tab:table2}
Formation enthalpy of different doping systems at different doping concentrations.}
\begin{tabular}{ccccc}
\hline
Dopant ($M$) & Doping concentration & $\Delta H_{f}(M_{Cu})$ (eV) & $\Delta H_{f}(M_{Al})$ (eV) & $\Delta H_{f}(M_{Se})$ (eV) \\ \hline
$M=$ Ti      & 25\%                 & 1.21  & 0.30  & 2.85 \\
             & 12.5\%               & 1.25  & 0.32  & 3.27 \\
             & 6.25\%               & 1.31  & 0.34  & 5.21 \\
             & 3.125\%              & 1.32  & 0.36  & 5.96 \\ \hline
$M=$ V       & 25\%                 & 1.72  & 1.24  & 4.13 \\
             & 12.5\%               & 1.77  & 1.24  & 4.44 \\
             & 6.25\%               & 1.82  & 1.25  & 5.75 \\
             & 3.125\%              & 1.83  & 1.25  & 5.77 \\ \hline
$M=$ Fe      & 25\%                 & 0.81  & 1.56  & 4.40 \\
             & 12.5\%               & 0.82  & 1.56  & 4.48 \\
             & 6.25\%               & 0.82  & 1.57  & 4.51 \\
             & 3.125\%              & 0.83  & 1.59  & 4.66 \\ \hline
\end{tabular}
\end{table}

\begin{figure}
\includegraphics[width=\columnwidth]{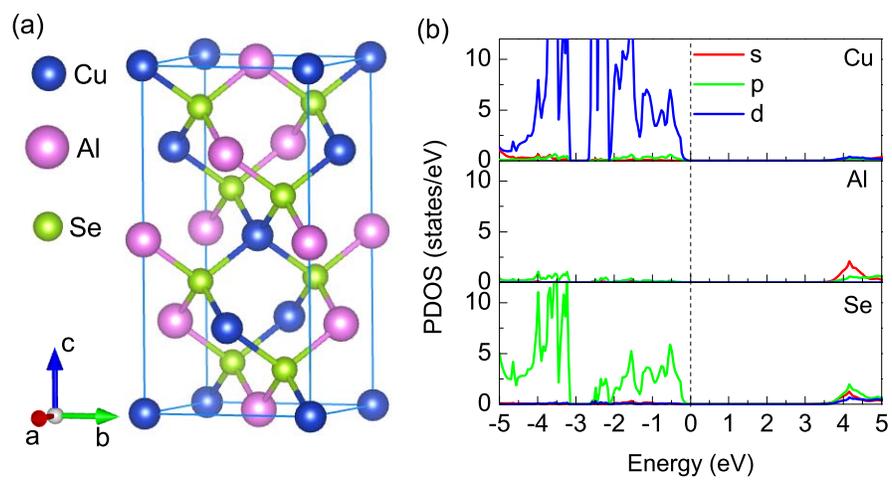}
\caption{(Color online) (a) Crystal structure and (b) projected density of states of CuAlSe$_2$.}
\end{figure}

\begin{figure*}
\includegraphics[width=\textwidth]{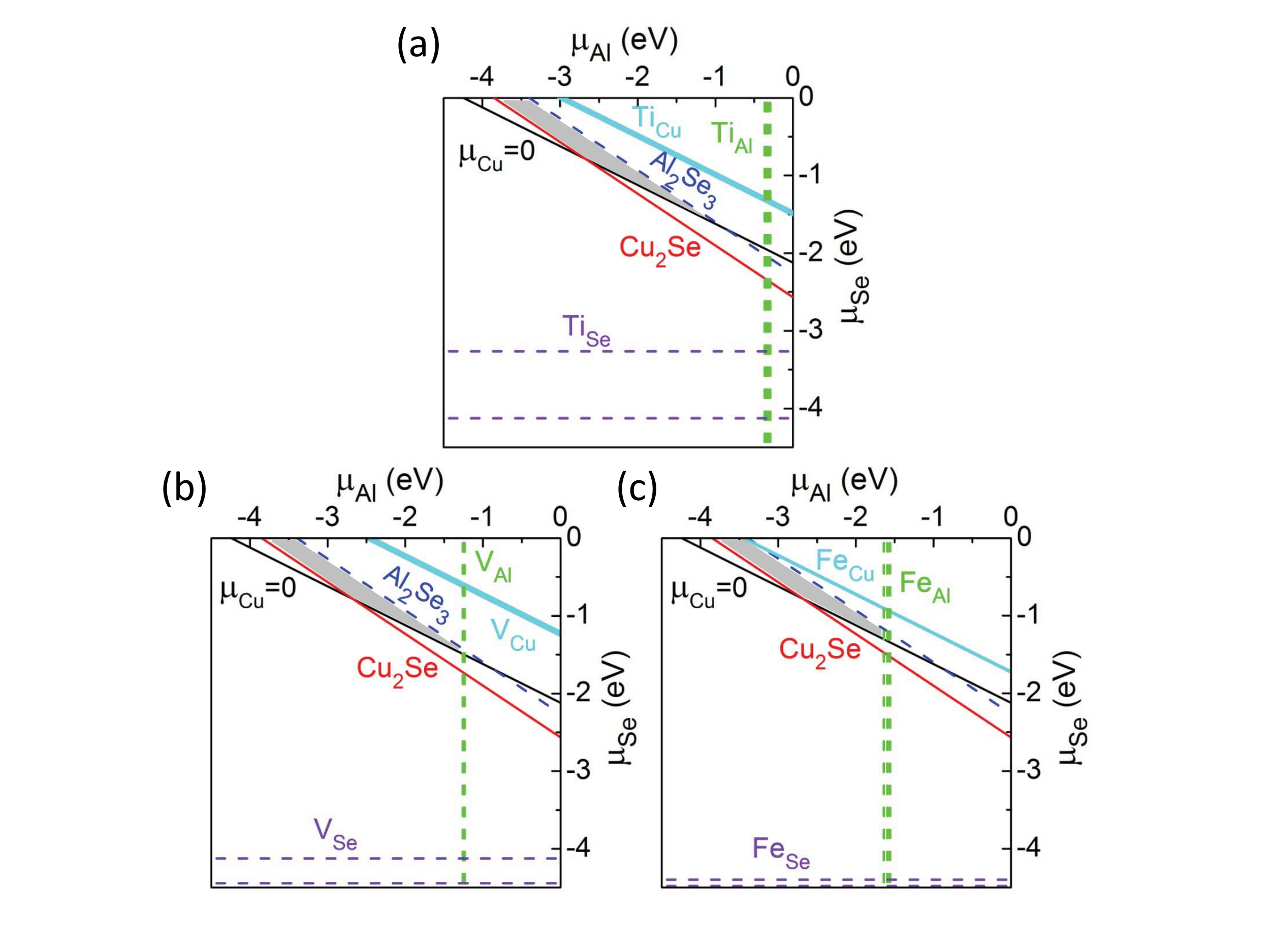}
\caption{(Color online) Calculated stable chemical potential range (the grey region) of doped CuAlSe$_2$. (a)-(c) are corresponding to Ti, V, and Fe  doping situations, respectively. For all chemical potential diagrams, the solid lines are used when the origin satisfies the constraints, and dashed lines are used when the origin does not satisfy the constraints. The black solid line indicates the position with $\mu_{Cu} = 0$, and the bed solid (blue dashed) line indicates the phase boundary of Cu$_2$Se (Al$_2$Se$_3$). For doping, the light blue, green dashed, and violet dashed line represent the chemical potential boundary of substituting $M$ atom for Cu, Al, and Se atom, respectively. The shift of the line determined by $M_{Cu}$, $M_{Al}$, or $M_{Se}$ implies the variation of doping concentration.}
\end{figure*}

\begin{figure}
\includegraphics[width=\columnwidth]{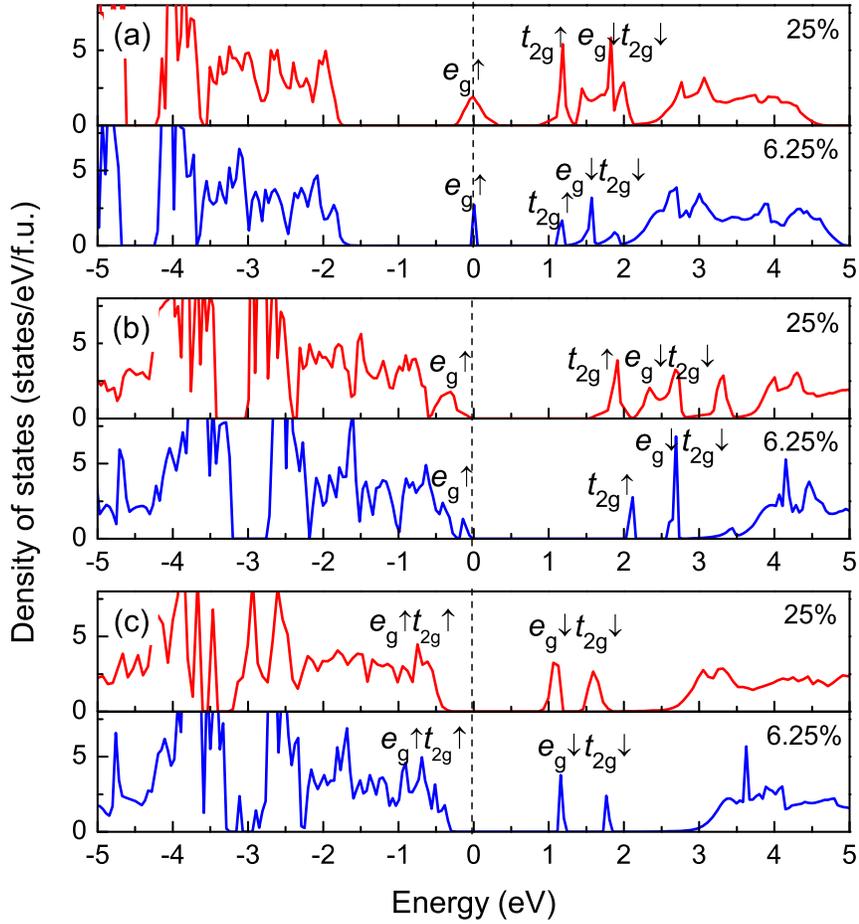}
\caption{(Color online) Calculated electronic density of states of doped CuAlSe$_2$. (a)-(c) are corresponding to Ti, V, and Fe doping cases, respectively. For each other element doping, two concentrations of 25\% (red line) and 6.25\% (blue line) are presented. The $3d$ electronic distributions of dopant also marked, where the arrow implies the direction of the electronic spin.}
\end{figure}

\begin{figure}
\includegraphics[width=\columnwidth]{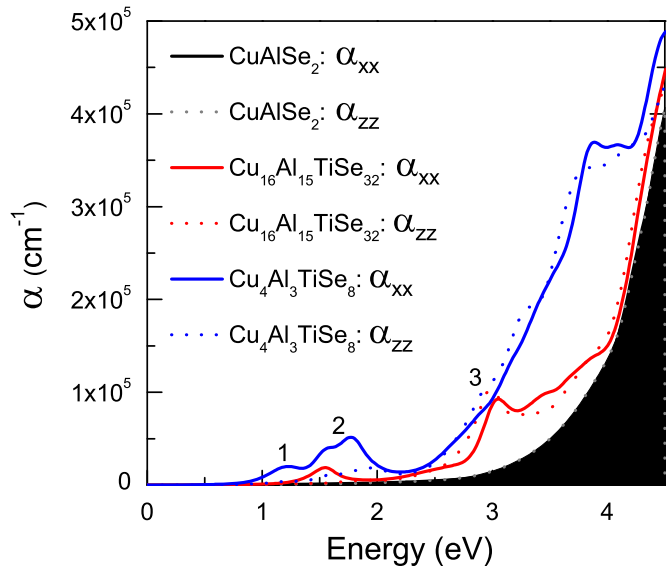}
\caption{(Color online) Calculated absorption coefficient $\alpha$ in visible light region of Ti-doped materials comparing with pure CualSe$_2$. The absorption coefficient is divided into the transverse contribution (xx; solid lines) and longitudinal contribution (zz; dotted lines).}
\end{figure}

\end{document}